\documentclass[longauth,letter,traditabstract]{aa}
\usepackage{graphicx}
\usepackage{txfonts}
\usepackage{natbib}
\bibpunct{(}{)}{;}{a}{}{,}
\begin{document}
\title{HerMES: The Submillimeter Spectral Energy Distributions of 
Herschel/SPIRE-Detected Galaxies\thanks{{\it Herschel} is an ESA space observatory with science
instruments provided by European-led Principal Investigator consortia and with 
important participation from NASA.}}

\author{B.~Schulz\inst{1,2}
\and C.P.~Pearson\inst{3,4}
\and D.L.~Clements\inst{5}
\and B.~Altieri\inst{6}
\and A.~Amblard\inst{7}
\and V.~Arumugam\inst{8}
\and R.~Auld\inst{9}
\and H.~Aussel\inst{10}
\and T.~Babbedge\inst{5}
\and A.~Blain\inst{1}
\and J.~Bock\inst{1,11}
\and A.~Boselli\inst{12}
\and V.~Buat\inst{12}
\and D.~Burgarella\inst{12}
\and N.~Castro-Rodr{\'\i}guez\inst{13,14}
\and A.~Cava\inst{13,14}
\and P.~Chanial\inst{5}
\and A.~Conley\inst{15}
\and L.~Conversi\inst{6}
\and A.~Cooray\inst{7,1}
\and C.D.~Dowell\inst{1,11}
\and E.~Dwek\inst{16}
\and S.~Eales\inst{9}
\and D.~Elbaz\inst{10}
\and M.~Fox\inst{5}
\and A.~Franceschini\inst{17}
\and W.~Gear\inst{9}
\and E.~Giovannoli\inst{12}
\and J.~Glenn\inst{15}
\and M.~Griffin\inst{9}
\and M.~Halpern\inst{18}
\and E.~Hatziminaoglou\inst{19}
\and E.~Ibar\inst{20}
\and K.~Isaak\inst{9}
\and R.J.~Ivison\inst{20,8}
\and G.~Lagache\inst{21}
\and L.~Levenson\inst{1,11}
\and N.~Lu\inst{1,2}
\and S.~Madden\inst{10}
\and B.~Maffei\inst{22}
\and G.~Mainetti\inst{17}
\and L.~Marchetti\inst{17}
\and G.~Marsden\inst{18}
\and A.M.J.~Mortier\inst{5}
\and H.T.~Nguyen\inst{11,1}
\and B.~O'Halloran\inst{5}
\and S.J.~Oliver\inst{23}
\and A.~Omont\inst{24}
\and M.J.~Page\inst{25}
\and P.~Panuzzo\inst{10}
\and A.~Papageorgiou\inst{9}
\and I.~P{\'e}rez-Fournon\inst{13,14}
\and M.~Pohlen\inst{9}
\and N.~Rangwala\inst{15}
\and J.I.~Rawlings\inst{25}
\and G.~Raymond\inst{9}
\and D.~Rigopoulou\inst{3,26}
\and D.~Rizzo\inst{5}
\and I.G.~Roseboom\inst{23}
\and M.~Rowan-Robinson\inst{5}
\and M.~S\'anchez Portal\inst{6}
\and Douglas~Scott\inst{18}
\and N.~Seymour\inst{25}
\and D.L.~Shupe\inst{1,2}
\and A.J.~Smith\inst{23}
\and J.A.~Stevens\inst{27}
\and M.~Symeonidis\inst{25}
\and M.~Trichas\inst{5}
\and K.E.~Tugwell\inst{25}
\and M.~Vaccari\inst{17}
\and E.~Valiante\inst{18}
\and I.~Valtchanov\inst{6}
\and L.~Vigroux\inst{24}
\and L.~Wang\inst{23}
\and R.~Ward\inst{23}
\and G.~Wright\inst{20}
\and C.K.~Xu\inst{1,2}
\and M.~Zemcov\inst{1,11}}

\institute{California Institute of Technology, 1200 E. California Blvd., Pasadena, CA 91125, USA\\
 \email{bschulz@ipac.caltech.edu}
\and Infrared Processing and Analysis Center, MS 100-22, California Institute of Technology, JPL, Pasadena, CA 91125, USA
\and Space Science \& Technology Department, Rutherford Appleton Laboratory, Chilton, Didcot, Oxfordshire OX11 0QX, UK
\and Institute for Space Imaging Science, University of Lethbridge, Lethbridge, Alberta, T1K 3M4, Canada
\and Astrophysics Group, Imperial College London, Blackett Laboratory, Prince Consort Road, London SW7 2AZ, UK
\and Herschel Science Centre, European Space Astronomy Centre, Villanueva de la Ca\~nada, 28691 Madrid, Spain
\and Dept. of Physics \& Astronomy, University of California, Irvine, CA 92697, USA
\and Institute for Astronomy, University of Edinburgh, Royal Observatory, Blackford Hill, Edinburgh EH9 3HJ, UK
\and Cardiff School of Physics and Astronomy, Cardiff University, Queens Buildings, The Parade, Cardiff CF24 3AA, UK
\and Laboratoire AIM-Paris-Saclay, CEA/DSM/Irfu - CNRS - Universit\'e Paris Diderot, CE-Saclay, pt courrier 131, F-91191 Gif-sur-Yvette, France
\and Jet Propulsion Laboratory, 4800 Oak Grove Drive, Pasadena, CA 91109, USA
\and Laboratoire d'Astrophysique de Marseille, OAMP, Universit\'e Aix-marseille, CNRS, 38 rue Fr\'ed\'eric Joliot-Curie, 13388 Marseille cedex 13, France
\and Instituto de Astrof{\'\i}sica de Canarias (IAC), E-38200 La Laguna, Tenerife, Spain
\and Departamento de Astrof{\'\i}sica, Universidad de La Laguna (ULL), E-38205 La Laguna, Tenerife, Spain
\and Dept. of Astrophysical and Planetary Sciences, CASA 389-UCB, University of Colorado, Boulder, CO 80309, USA
\and Observational  Cosmology Lab, Code 665, NASA Goddard Space Flight  Center, Greenbelt, MD 20771, USA
\and Dipartimento di Astronomia, Universit\`{a} di Padova, vicolo Osservatorio, 3, 35122 Padova, Italy
\and Department of Physics \& Astronomy, University of British Columbia, 6224 Agricultural Road, Vancouver, BC V6T~1Z1, Canada
\and ESO, Karl-Schwarzschild-Str. 2, 85748 Garching bei M\"unchen, Germany
\and UK Astronomy Technology Centre, Royal Observatory, Blackford Hill, Edinburgh EH9 3HJ, UK
\and Institut d'Astrophysique Spatiale (IAS), b\^atiment 121, Universit\'e Paris-Sud 11 and CNRS (UMR 8617), 91405 Orsay, France
\and School of Physics and Astronomy, The University of Manchester, Alan Turing Building, Oxford Road, Manchester M13 9PL, UK
\and Astronomy Centre, Dept. of Physics \& Astronomy, University of Sussex, Brighton BN1 9QH, UK
\and Institut d'Astrophysique de Paris, UMR 7095, CNRS, UPMC Univ. Paris 06, 98bis boulevard Arago, F-75014 Paris, France
\and Mullard Space Science Laboratory, University College London, Holmbury St. Mary, Dorking, Surrey RH5 6NT, UK
\and Astrophysics, Oxford University, Keble Road, Oxford OX1 3RH, UK
\and Centre for Astrophysics Research, University of Hertfordshire, College Lane, Hatfield, Hertfordshire AL10 9AB, UK}

\date{Received Mar 31, 2010; accepted May 13, 2010}

\abstract
{We present colours of sources detected with the {\em Herschel/SPIRE} instrument in deep 
extragalactic surveys of the Lockman Hole, Spitzer-FLS,
and GOODS-N fields in three photometric 
bands at 250, 350 and 500~$\mu$m. We compare these with expectations
from the literature and discuss associated uncertainties and biases in the SPIRE data.
We identify a $500$~$\mu$m flux limited
selection of sources from the HerMES point source 
catalogue that appears free from neighbouring/blended sources in all three SPIRE bands.
We compare the colours with redshift tracks of various contemporary models.
Based on these spectral templates we show that regions corresponding to specific population 
types and redshifts can be identified better in colour-flux space.
The redshift tracks as well as the colour-flux plots imply a majority of detected objects 
with redshifts at 1$<$z$<$3.5, somewhat depending on the group of model SEDs used.
We also find that a population of S$_{250}$/S$_{350}<0.8$ at fluxes above 50~mJy as observed 
by SPIRE is not well represented by contemporary models and could consist of a mix of 
cold and lensed galaxies.
}


\maketitle

%

\section{Introduction}\label{sec:introduction}

Galaxy formation and evolution are major topics in cosmology and astrophysics. 
Recently, sophisticated computer simulations based on hierarchical formation
(eg. \citep{springel05}) and deep cosmological surveys across the electromagnetic
spectrum have transformed the field from a collection of a few hypothetical models 
to a mature science that is profoundly shaping our understanding of the universe.  
The Spectral and Photometric Imaging Receiver (SPIRE, \citep{griffin10}), on board of 
ESA's {\it Herschel} observatory \citep{pilbratt10}, opens a new window between 
250 -- 500~$\mu m$ for observations of heavily dust obscured high redshift 
($z \sim 2$ -- 3) galaxies.
The combination of sensitivity and high resolution provided by its 3.5m telescope, 
the largest ever launched, allows large area confusion limited far-infrared and 
submillimeter (FIR/submm) surveys to reach unprecedented depths, exploring previously 
undetectable remote galaxy populations. Studies of the integrated cosmic 
background light have shown that at least half the radiation from all galaxies lies 
in the FIR/submm. In particular, galaxies at $z \geq 1$ radiate mostly in this band
\citet{lagache05}. The peak of the spectral energy distribution (SED) from these
galaxies is redshifted into the SPIRE bands. Deep SPIRE surveys will thus provide 
a new census of the energy budget in $z \geq 1$ galaxies, especially for the most 
energetic dust-rich objects.

In this paper we investigate SPIRE colours in three bands, 250~$\mu m$, 350~$\mu m$, 
and 500~$\mu m$, of a sample of sources selected at 500~$\mu m$ in four Science Demonstration
Phase (SDP) fields of the Herschel Multi-tired Extragalactic Survey 
(HerMES\footnote{hermes.sussex.ac.uk}) key project \citep{oliver10}.
Our goal is to constrain the evolution of the IR SED of FIR/submm galaxies
(i.e. the SED vs. redshift relation).  In particular we want to examine
whether there is a new population of submm galaxies whose SEDs do not match
the predictions of current models.

\smallskip

\section{Observations and data processing}\label{sec:observations}
The HerMES key project is constructed in order to obtain a complete bolometric census of 
star-formation in the Universe. It consists of 6 tiers of survey fields with increasing 
depth over smaller areas, covering most of the 
fields on the sky observed across the
electromagnetic spectrum by state-of-the-art facilities
plus individual selected clusters. 
A total of 4 HerMES fields were surveyed during 
{\it Herschel's} SDP and we have used the deep observations in 
GOODS-N, Lockman-North, FLS, and Lockman-SWIRE for our analysis. The covered areas 
are 0.25, 0.34, 5.81 and 13.2 deg$^2$ 
respectively with relative depths of 1.0, 0.23, 0.05, 0.033 that were calculated as the fraction of 
the number of repeats and scan speed, normalised to the deepest field GOODS-N.
More details of the observations are given by \citet{oliver10}.

Data processing based on the standard SPIRE Scan Map Pipeline \citep{griffin08} yielded maps 
in the three SPIRE bands, and source catalogues for each individual band were generated using 
the SUSSEXtractor software \citep{savage07} within HIPE 3.0 \citep{ott06}.
The three shallower maps were smoothed with point-source optimised filters while the deepest 
map was filtered with a delta function to find sources and separately with a $3 \times 3$ pixel 
point source response function (PRF)
to extract the fluxes \citep{oliver10}. The FLS and Lockman-SWIRE fields were Wiener
filtered to reduce effects by diffuse Cirrus.
For the source extraction a Gaussian 
PRF 
was assumed, 
with FWHM of 18.2$^{\prime\prime}$, 25.2$^{\prime\prime}$  and 36.3$^{\prime\prime}$ for the 
SPIRE 250, 350 \& 500~$\mu$m bands, respectively.
Details of the procedure are given by \citet{oliver10} and \citet{smith10}.
They attained formal 1-$\sigma$ point source uncertainties of 
5.7, 7.4 and 7.8~mJy for GOODS-N, 
7.0, 8.5 and 8.8~mJy for Lockman-North,
9.0, 10.3 and 10.6~mJy for FLS, and
11.1, 16.9 and 15.1~mJy for Lockman-SWIRE,
respectively.
These numbers include a contribution from source confusion of approximately 
5.6, 7.4, and 7.7~mJy for GOODS-N, 6.8, 8.3 and 8.5~mJy for Lockman-North, 
8.4, 9.8, 9.7~mJy for FLS, and 8.5, 13.9, 10.2 for Lockman-SWIRE.
For the two deepest fields
\citet{smith10} attribute the differences to the 
results of \citet{nguyen10}, and the differences between the fields mainly to the 
source extraction method used.

The catalogues contain additional parameters to allow for quality checking and source selection. 
These are:
{\it i})~A formal error in the flux measurement, propagated through from the error maps created 
by the map-maker, representing a fair estimate of the instrumental noise.
{\it ii})~A total error that is the quadratic co-addition of instrument noise and average 
estimated confusion noise over the map.
{\it iii})~Two separate flux estimates for the same positions using two different halves 
of the data (half-maps), separated in time, allowing for the detection and exclusion of spurious 
sources, mostly due to high energy particle hits.

   \begin{figure}
   \centering
   \includegraphics[width=6.5cm]{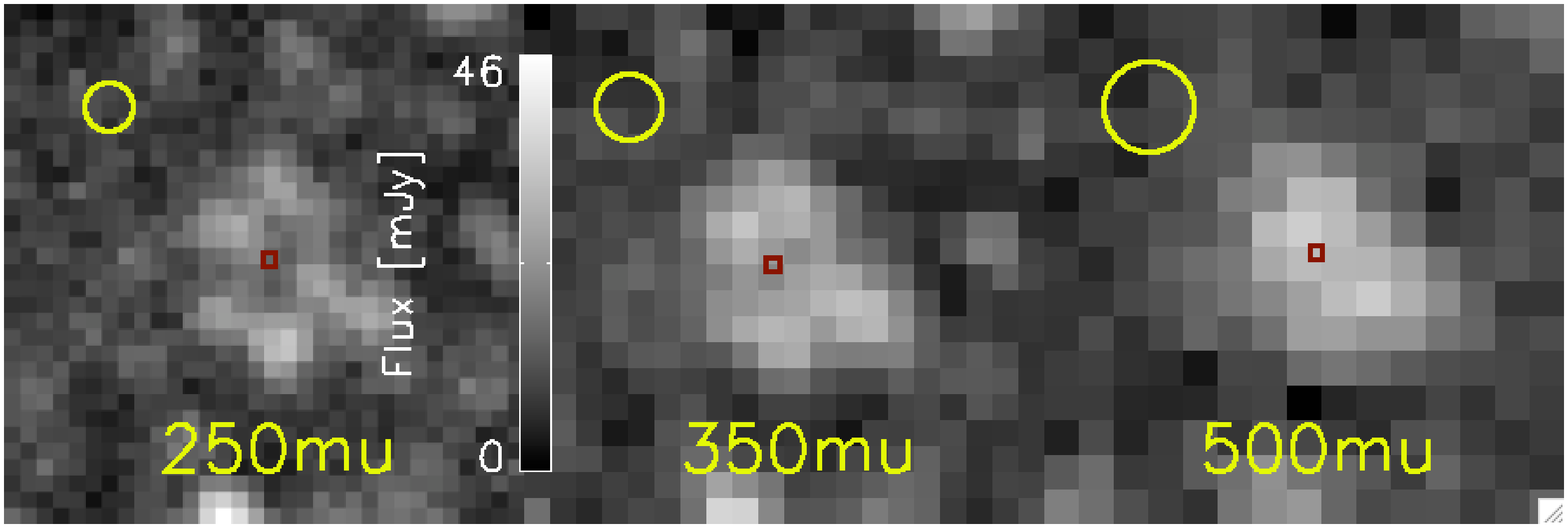}
   \includegraphics[width=6.5cm]{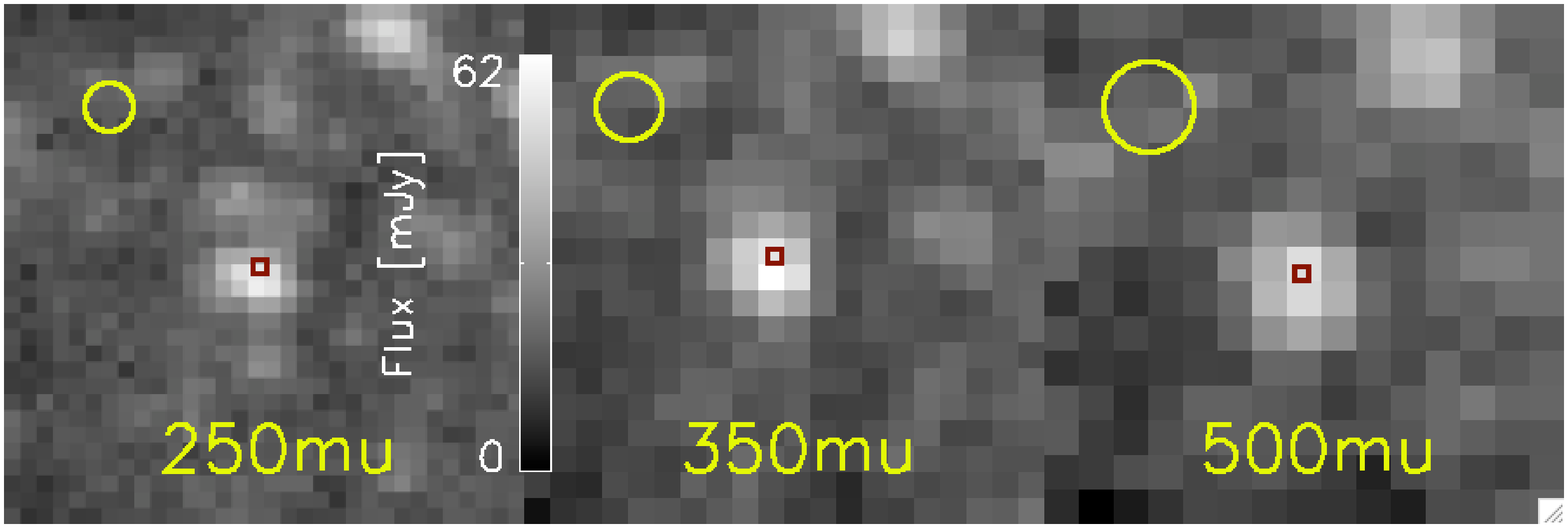}
      \caption{Examples of single 500~$\mu$m source detections ({\it top 
      and bottom panel, right}) 
      with multiple counterparts at 350~$\mu$m and 250~$\mu$m ({\it top panel, middle and left}), 
      and single 350 and 250~$\mu$m counterparts ({\it bottom panel, middle and left}).
      Note that the upper 500~$\mu$m source appears already double to the eye. 
      The red marked source position in each image is the one determined 
      by the algorithm 
      at 500~$\mu$m. Each image measures 3$^\prime$ on each side. 
      The yellow circles indicate the FWHM of the beams.
      }
         \label{Fig:multisource}
   \end{figure}

\smallskip

\section{Catalogue cross association}\label{sec:associations}

The starting point for the cross-association process is this set of individual SPIRE band 
catalogues. For the current work the emphasis is on the creation of a robust, un-confused 
sample of sources that has the highest probability for its colours to originate from single 
unblended galaxies.
We only consider the central regions of the maps, that have full homogeneous coverage by 
all scans.
To protect against spurious sources, we compare fluxes separately derived from two independent
half-maps. The ratio of the two flux estimates separates well into 3 distributions. 
Spurious sources are removed 
 by excluding ratios above 5 and below $1/5$.
For this work we have constructed a 500~$\mu$m band flux limited selection.
It is justified in three ways:
i) The stronger negative $K$-correction means selection in this band favours higher redshift galaxies; 
ii) This is a relatively new band as yet only explored by much shallower BLAST surveys \citep{devlin09};
iii) About $\sim 90$\% of 500~$\mu$m selected sources are also detected in the other SPIRE bands.
We require a signal-to-total-noise (S/N) ratio of more than 3 
 in the 500~$\mu$m filter.
The formal average flux uncertainties at 
 500~$\mu$m
 derived from the source extractor results 
are only 
 1.1, 2.1, 4.4, and 11.6~mJy.
We consider these to be instrumental noise, based on their ratio, consistent with the 30, 7, 
 2, and 1 
repetitions executed on the 
 four fields,
 respectively, 
  and the $1 \sigma$ confusion noise at 500~$\mu$m of 
6.8$\pm$0.4 reported by \citet{nguyen10}.
Thus the uncertainties are completely dominated by extragalactic confusion 
 for the first two fields
and increased for FLS and Lockman-SWIRE.
We calculate the following effective $3 \sigma$ average flux limits in our fields from 3 times the 
average total error of all sources with S/N$<4$: 
23.4, 26.4, 32.0, and 46.4~mJy for GOODS-N, Lockman-North, FLS, and Lockman-SWIRE respectively. 
The selection leaves 48, 61, 608, and 824 sources at 500~$\mu$m in the 4 bands respectively.
This conservative 
threshold also minimises the impact of flux boosting on the derived colours of the sources.

To further de-blend 
 and cross-match, first, all 500~$\mu$m sources without another 500~$\mu$m
 source within an 18$^{\prime\prime}$ radius 
are selected. 
This radius was chosen to be similar to the beam size at 500~$\mu$m. Then for these remaining 
sources, the same 18$^{\prime\prime}$ radius is checked in the other two bands. Sources with 
more than one source in a different band are discarded immediately. In case only one source 
is found in the other band, it must be within a radius of 8$^{\prime\prime}$ in order to be 
accepted as a cross identification, otherwise the source is considered blended and discarded. 
This radius was chosen to include 3-$\sigma$ of the telescope 
pointing error and the estimated PRF fit error of ~$6^{\prime\prime}$ each.
 We end up with a list of potentially uncontaminated 500~$\mu$m sources that is then cross
matched with the lists of the other two bands with a match radius of  8$^{\prime\prime}$. 
In Figure \ref{Fig:multisource} the dangers of simple naive associations are emphasised, 
where a cluster of sources shows up as being 
 detected as single at 500~$\mu$m by the point source extractor, 
but revealing multiple counterparts at 250~$\mu$m.
This sample is largely free from contamination and should have reliable fluxes originating 
from just one source, accurate at a 30~$\%$ level or better. 
 The final matched source numbers 
for the four fields respectively are 21, 38, 242, and 244.

\smallskip

\section{Analysis}\label{sec:analysis}

\begin{figure}
   \centering
   \includegraphics[width=4.6cm]{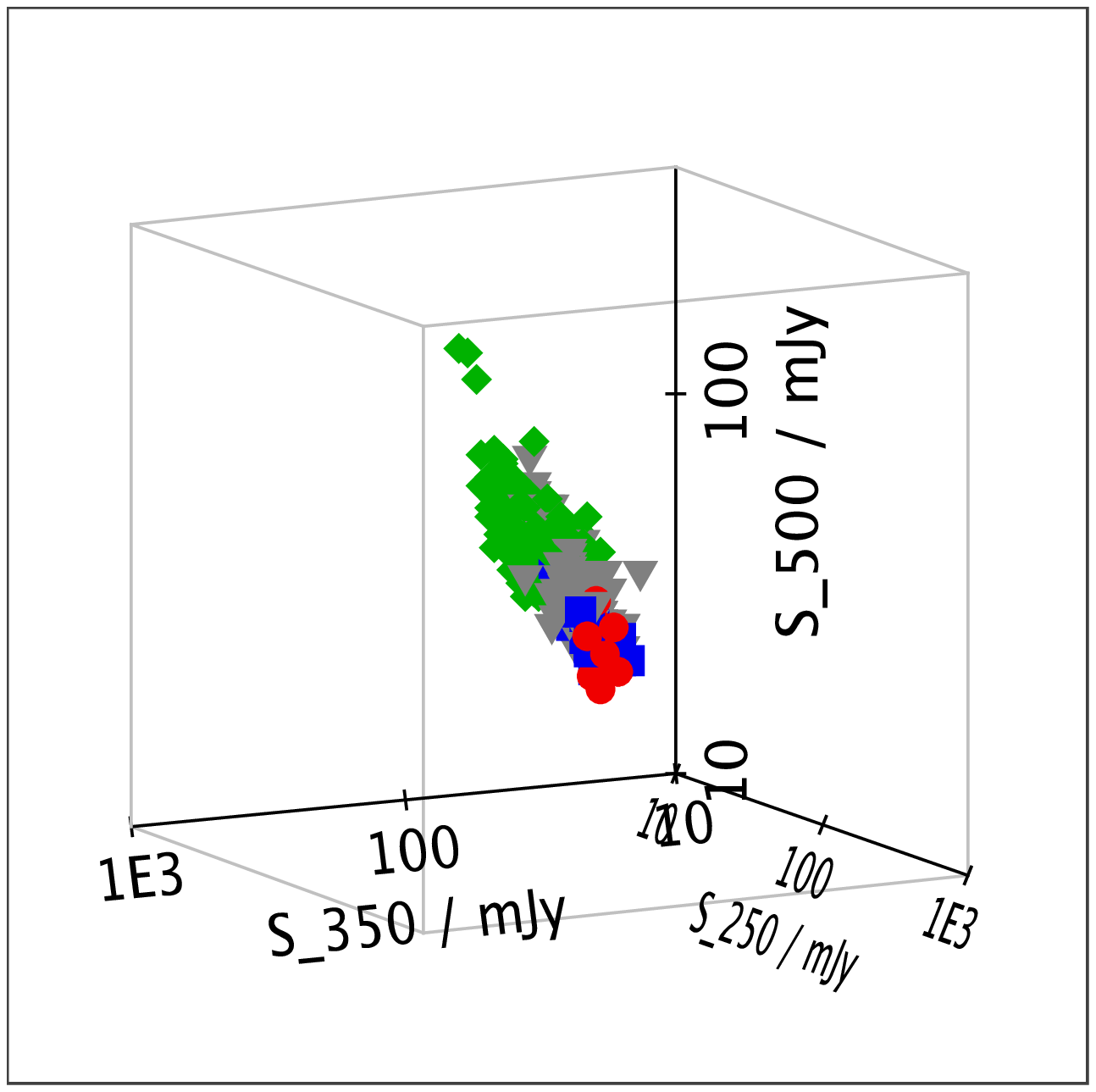}
   \includegraphics[width=4.3cm]{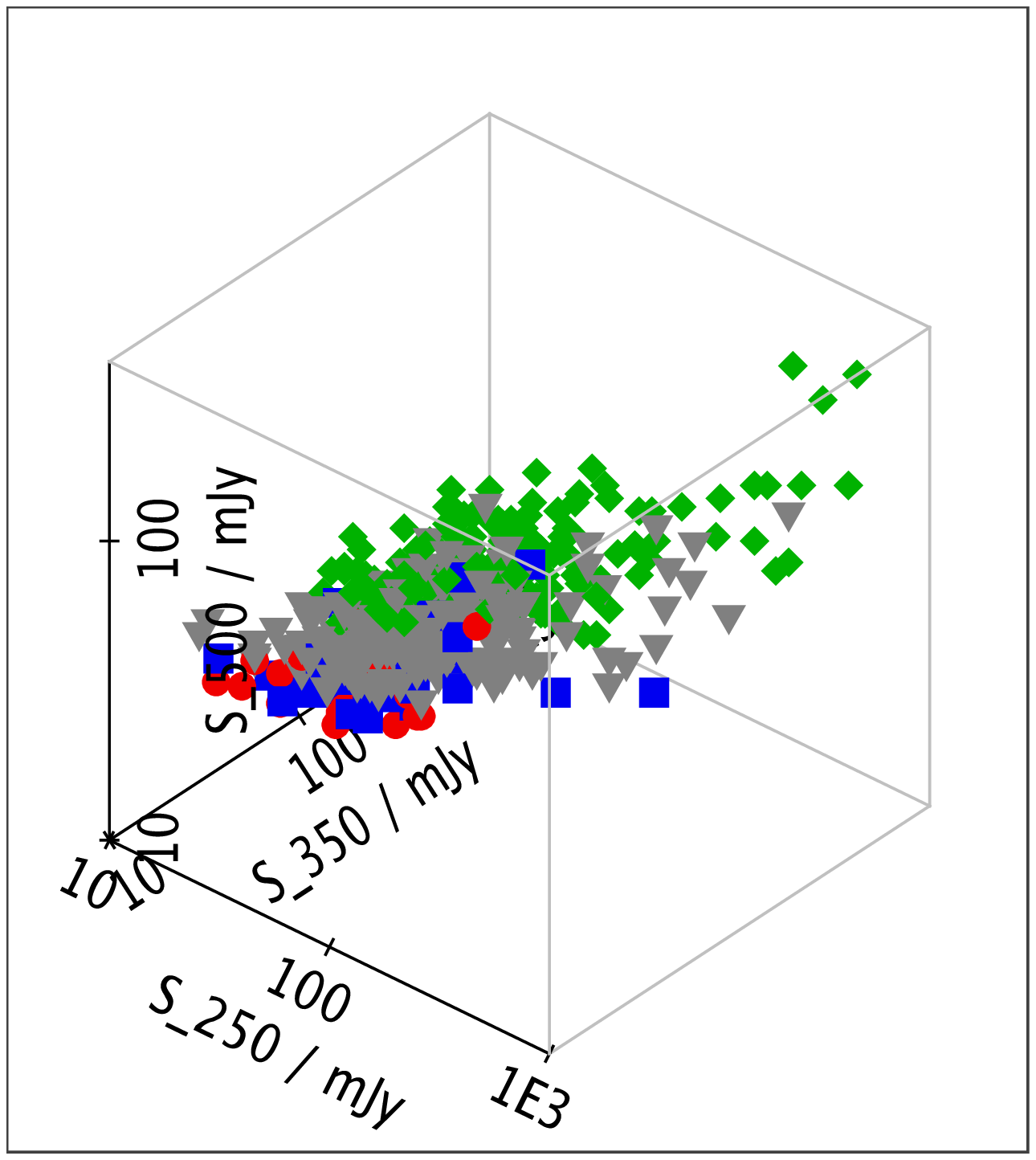}
      \caption{The 3-dimensional flux parameter space for our unblended band-merged 
      catalogues in the SPIRE 250, 350 \& 500~$\mu$m bands. The sources in 
      GOODS-N appear in red, Lockman-North in blue, 
      FLS in grey, and Lockman-SWIRE in green.
      Both diagrams show the same 3D plot from two different aspect angles, 
      the left one from
      within a plane fitted through the data, the other from a perpendicular direction.
      }
      \label{Fig:fluxfluxflux}
   \end{figure}

In Figure \ref{Fig:fluxfluxflux} we plot the 3-dimensional SPIRE flux-flux-flux parameter 
space for our band merged catalogue. The fluxes are grouped around a relatively flat and 
thin surface in the 250~$\mu$m, 350~$\mu$m, 500~$\mu$m parameter space. The same even thinner 
surface is seen in similar plots of mock catalogue data that is discussed later.
Thus, although we have flux data in three SPIRE bands, in principle only two parameters
are needed to describe the information. This degeneracy follows from the fact that the
spectral energy distributions (SEDs) in the submm, which SPIRE observes, are 
dominated by dust emission that have very similar shapes and result in fairly well defined 
flux ratios. Thus, the main parameters determining the three SPIRE fluxes, are rather 
wavelength of the emission peak and luminosity.

\smallskip

\subsection{Colour - colour  parameter space and model comparison}\label{sec:colour-colour}

\begin{figure*}
  \centering
  \centerline{
    \includegraphics[width=6.4cm]{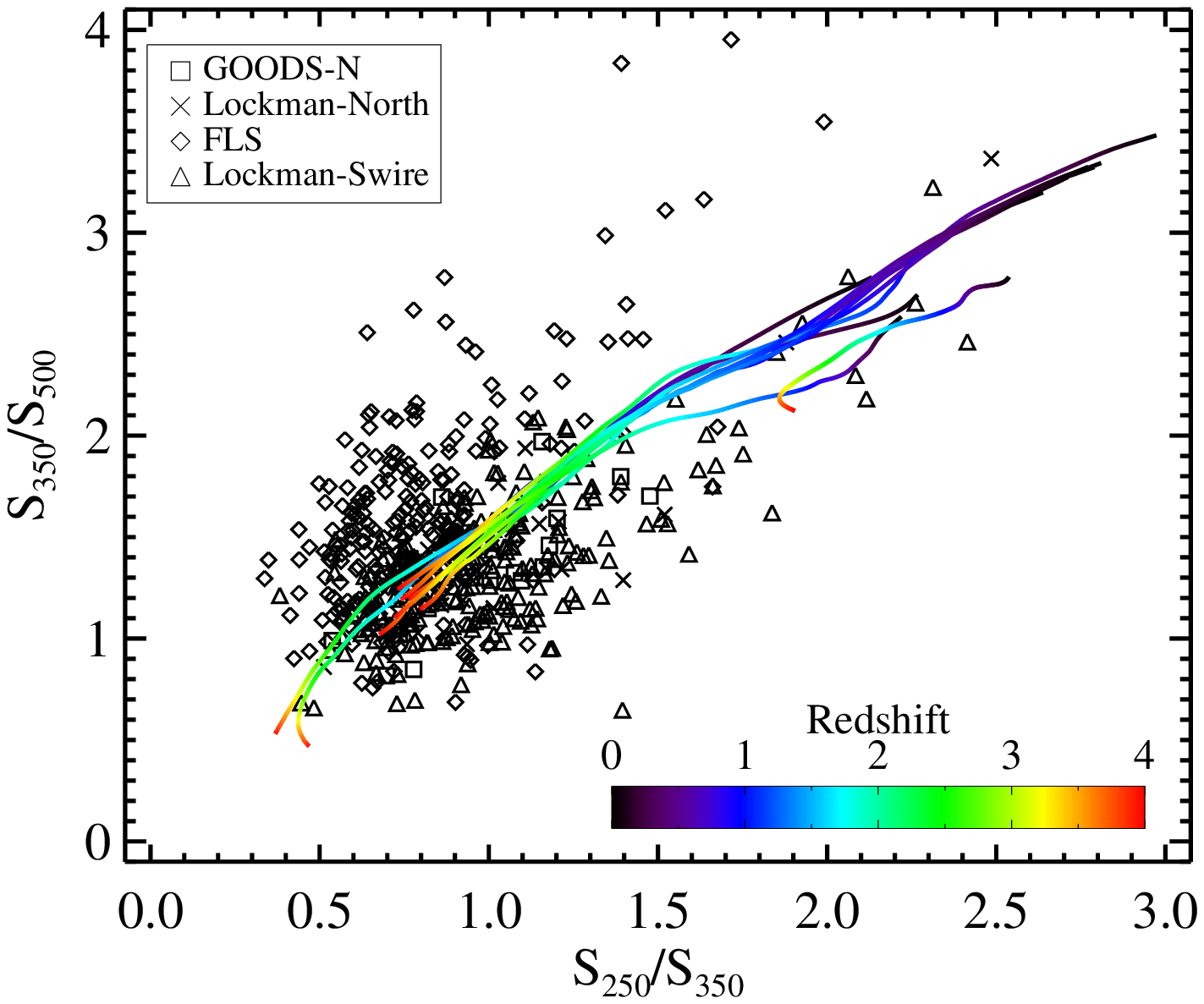}
    \includegraphics[width=6.4cm]{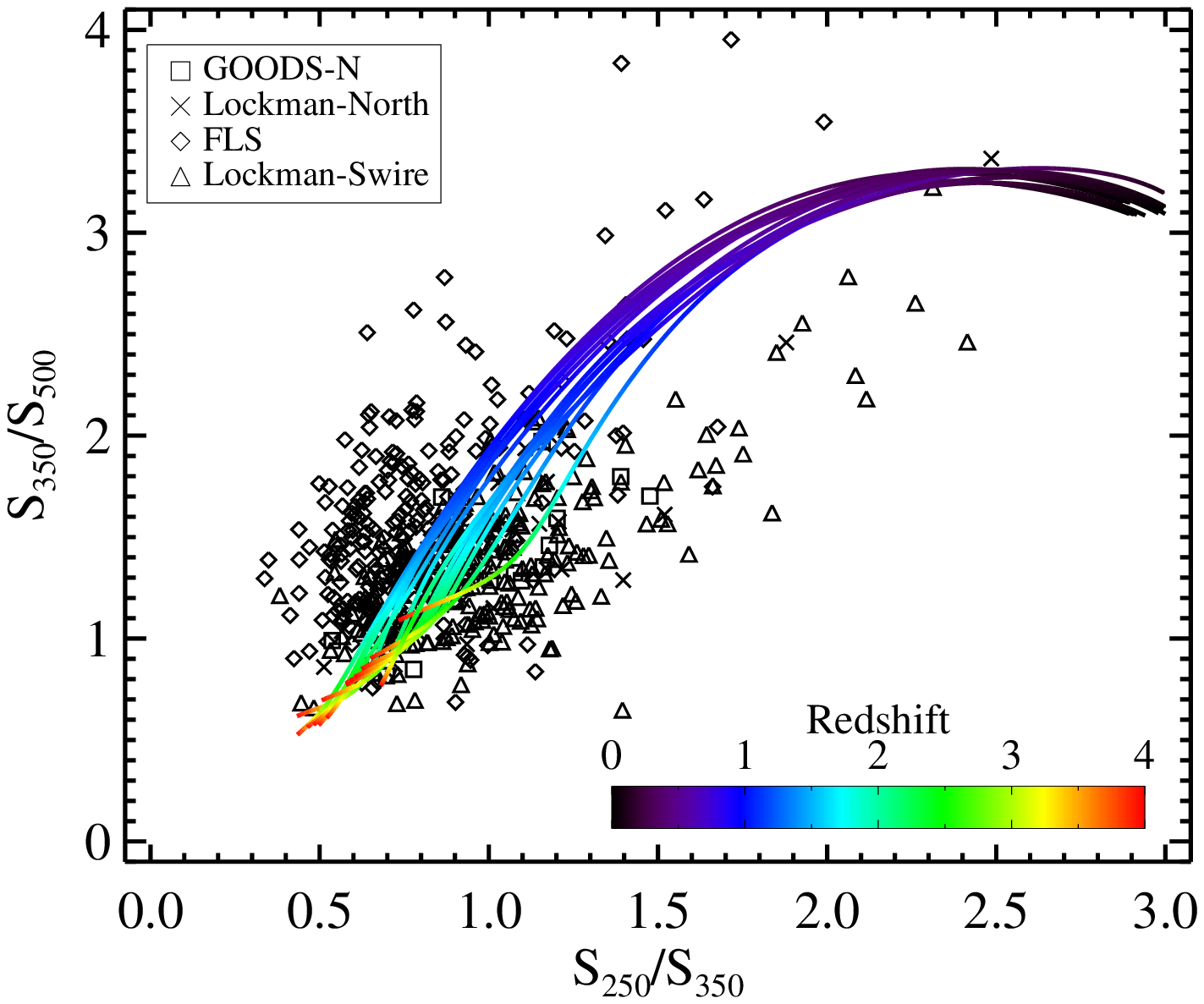}
  }
  \vspace{0.2cm}
  \centerline{
    \includegraphics[width=6.4cm]{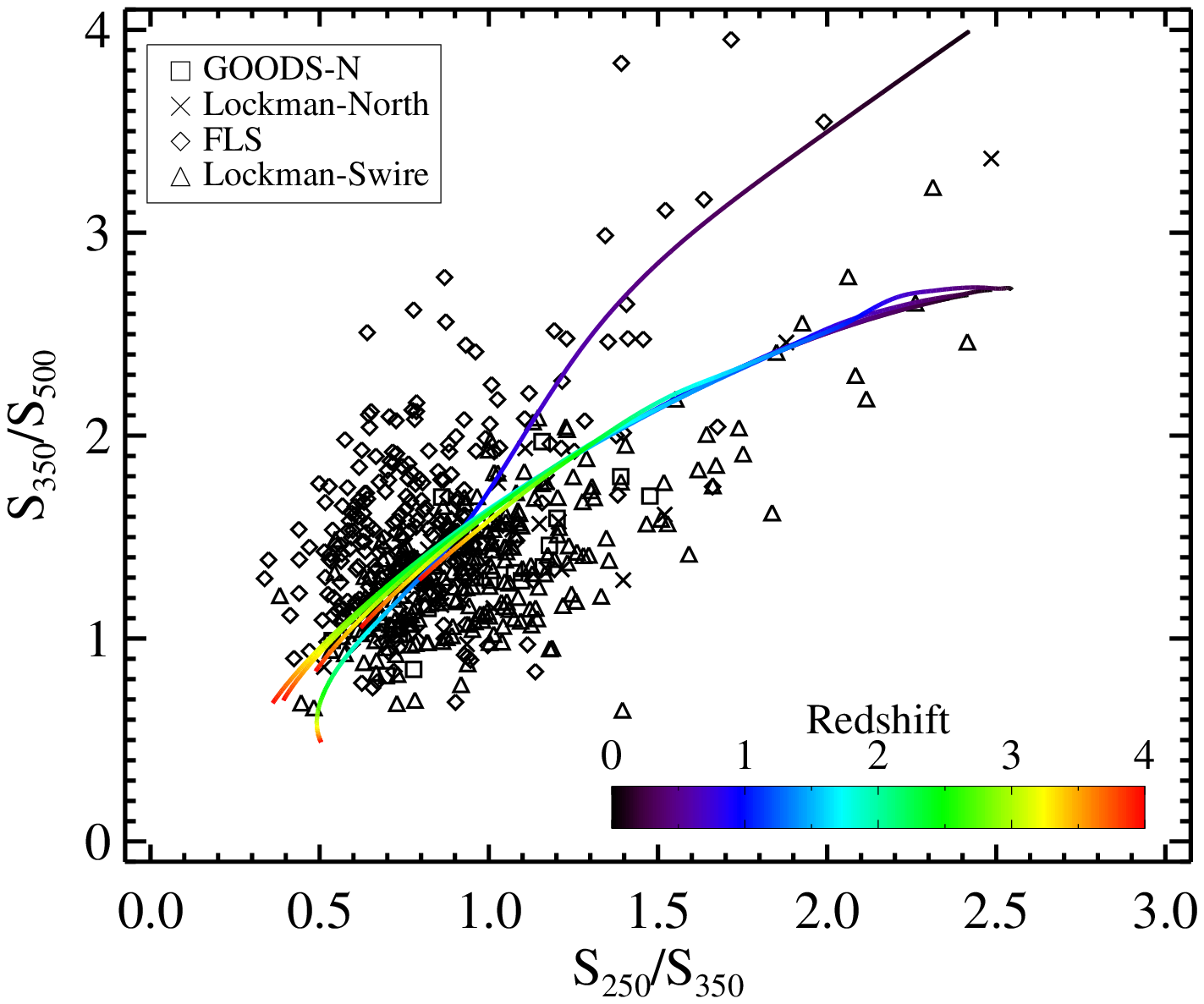}
    \includegraphics[width=6.4cm]{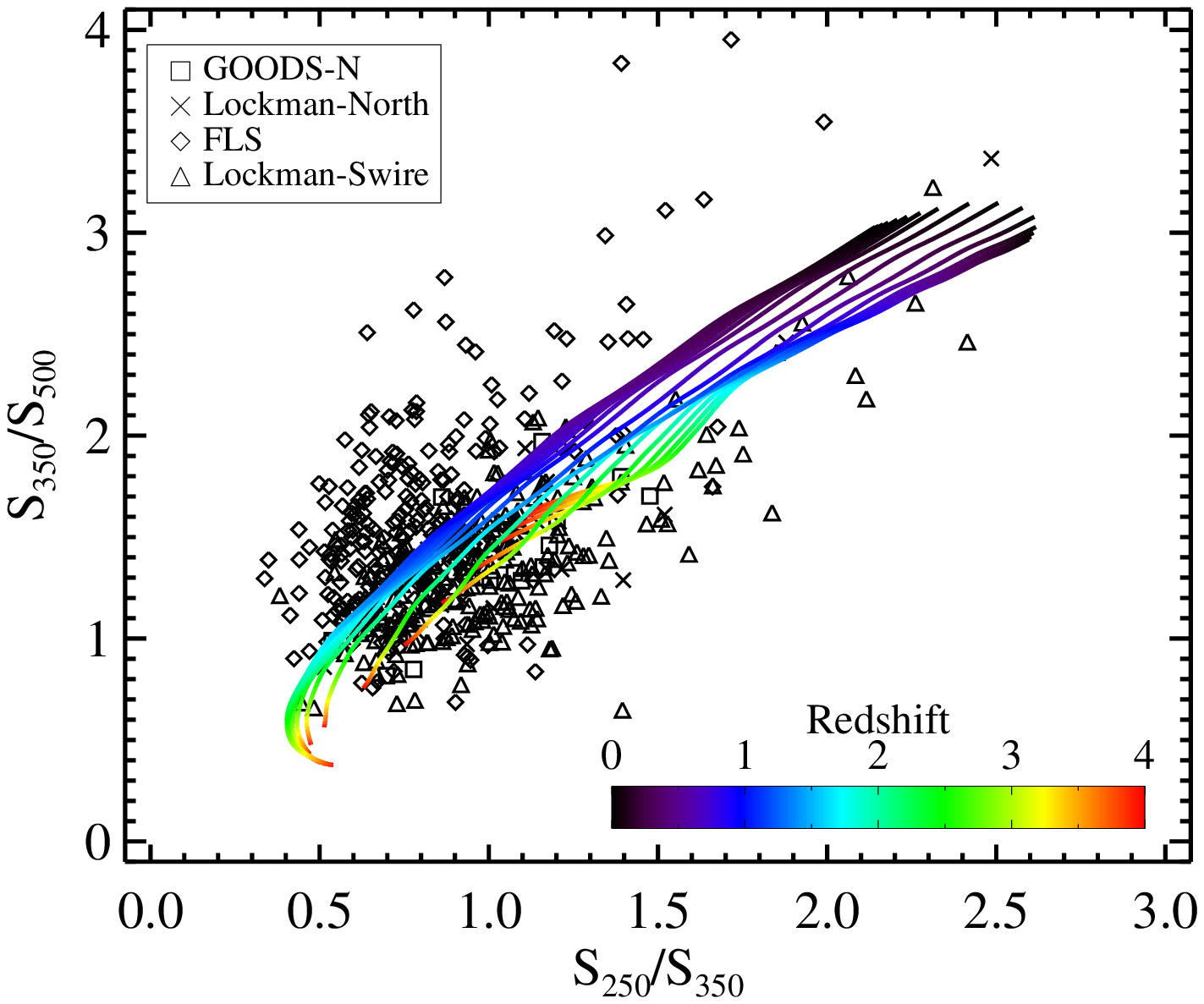}
  }
  \caption{ 
    S$_{350}$/S$_{500}$ - S$_{250}$/S$_{350}$ colour-colour plots for the SPIRE sources. 
    Over plotted are the colour tracks from the galaxy evolution models of 
    {\it top-left}  \citet{pearson07},
    {\it top-right} \citet{xu01},  
    {\it bottom-left} \citet{lagache03}, and 
    {\it bottom-right} \citet{dale02}. 
    The redshift in the tracks is colour coded and runs from 0 to 4. 
    The black symbols represent all unblended SPIRE sources according to the legend.
  }
  \label{Fig:colourcolour}
\end{figure*}

In the Figure~\ref{Fig:colourcolour} the S$_{250}$/S$_{350}$ - S$_{350}$/S$_{500}$ 
colour-colour diagrams for SPIRE sources in the SDP survey fields are shown with the colour 
tracks from the contemporary galaxy evolution models of \citet{pearson07}, \citet{dale02}, 
\citet{xu01} \&  \citet{lagache03} over-plotted on individual panels.
The redshift of the tracks is 
shown 
in colour and ranges from 0 to 4. 

In general, all the models are consistent with the obtained SPIRE colours, except from 
individual subtleties of the models. 
However, especially the SED templates of \citet{pearson07} and \citet{lagache03} lack diversity 
in dust illumination conditions to cover the spread of colours sufficiently.
All models agree: the colours imply that the SPIRE population is not local but rather the 
bulk lies at redshifts between 1 and 3.5. We note that the \citet{xu01} and \citet{dale02} 
models tend to place the population at somewhat lower redshift than the other two. 
 This implies that the \citet{pearson07} and \citet{lagache03} model SEDs contain generally 
warmer dust, which is confirmed by plots of the emission maxima of the SEDs.

\smallskip

\subsection{Colour - flux  parameter space}\label{sec:colour-flux}

\begin{figure*}
\centering
\centerline{
 \includegraphics[width=6.4cm]{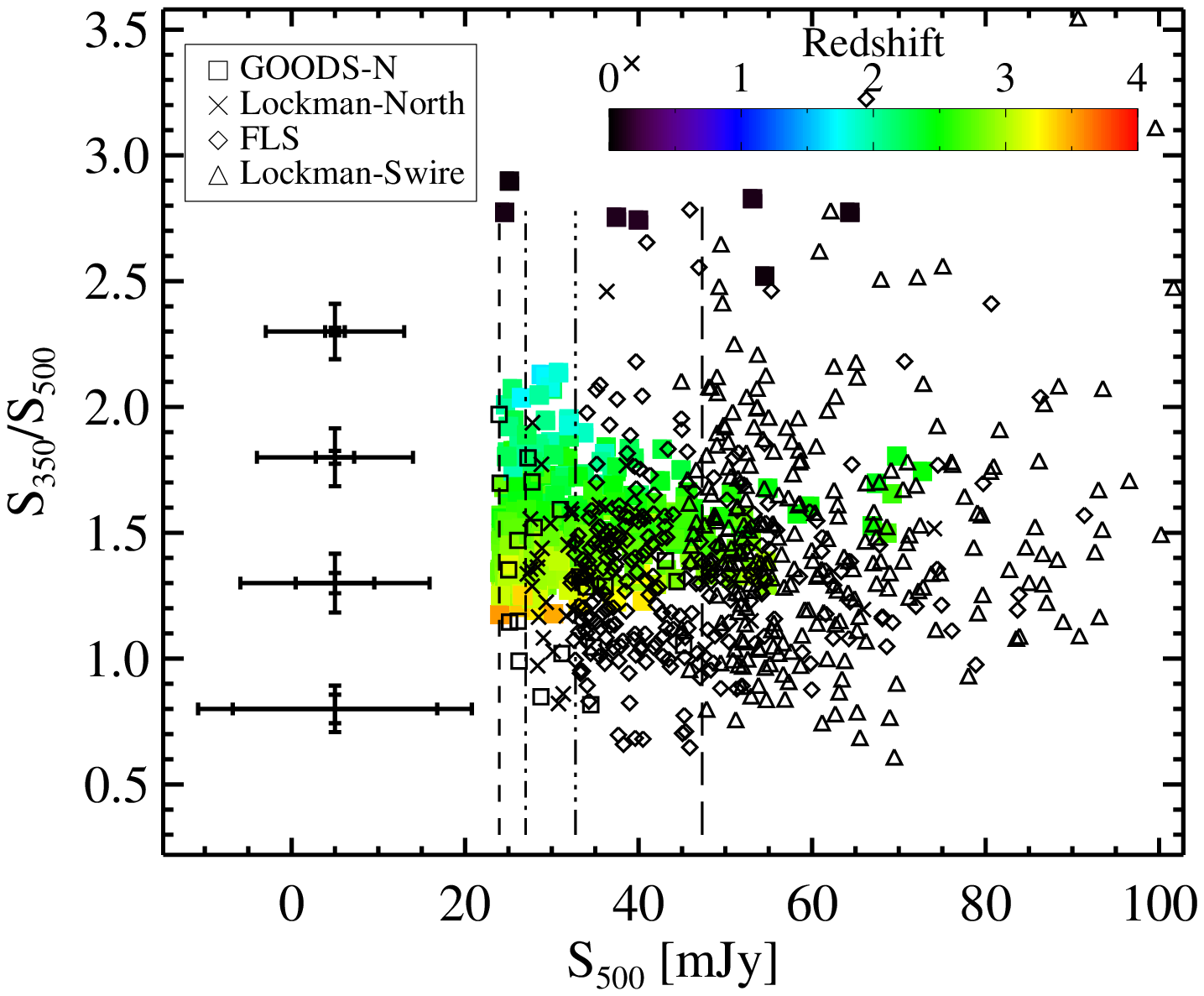}
 \includegraphics[width=6.4cm]{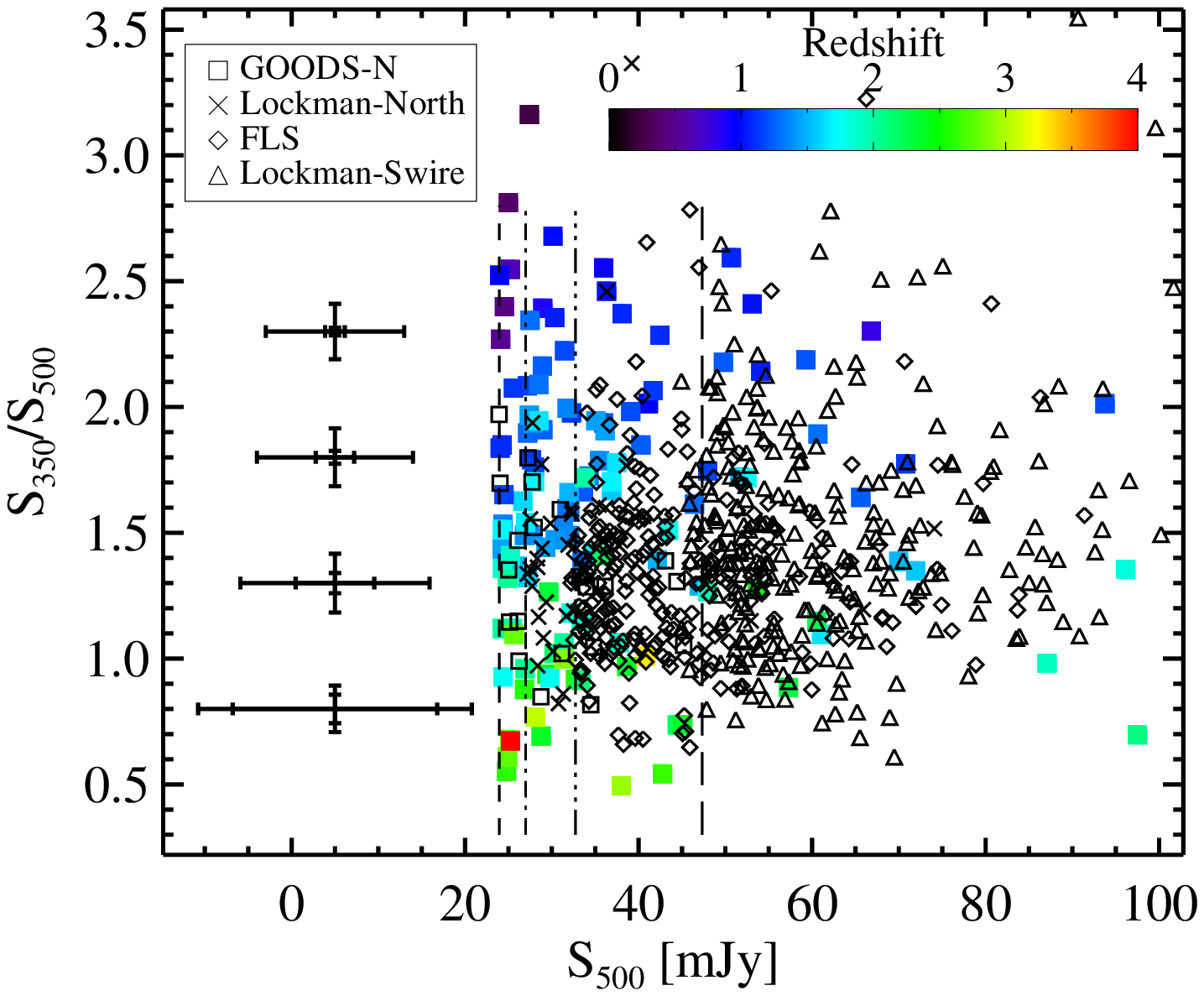}
}
\caption{Measured  
S$_{250}$/S$_{350}$ colour 500~$\mu$m
flux distributions for the SPIRE 
sources (black symbols) in comparison with mock catalogues of \citet{pearson07} to the {\it left} 
and \citet{xu01} on the {\it right}. The large error crosses on the left
represent average 1-$\sigma$ 
total uncertainties dominated by extragalactic confusion and the smaller tick marks show 
instrumental noise only, which is negligible for GOODS-N and Lockman-N.
The four vertical lines indicate from left to right,
the effective flux limits of GOODS-N, Lockman-North, FLS, and Lockman-SWIRE respectively.
}
\label{Fig:colourflux}
\end{figure*}

To overcome the apparent degeneracy in the SPIRE colours (Figure~\ref{Fig:fluxfluxflux}), 
we plot colour-flux distributions.
In Figure \ref{Fig:colourflux} we show the 
 S$_{250}$/S$_{350}$ colour versus 500~$\mu$m 
flux density distributions for the SPIRE sources. 
 A few sources at S$_{500}$$>$$100$~mJy are 
not shown to improve visibility. 
The symbols indicate the different fields according to the legend in the upper left corner.
The four crosses on the left are in the same vertical order as the symbols in the legend
and represent the averaged uncertainties in the four fields. Different tick marks show 
instrumental and total components.
The 
four
vertical lines indicate from left to right,
the effective flux limits of GOODS-N, Lockman-North, FLS, and Lockman-SWIRE respectively.
In both panels of Figure \ref{Fig:colourflux} the observed data are compared to mock catalogues of 1 deg$^2$
on the sky by \citet{pearson07} ({\it left}) and \citet{xu01} ({\it right}) that were cut 
below 
the effective flux limit of GOODS-N.
Again the most notable difference is the larger spread of the \citet{xu01} colours 
due to a larger number and diversity of SED models.
In both models the bulk of objects are Starburst galaxies, LIRGs and ULIRGs, that are grouped 
around a colour of S$_{250}$/S$_{350}\approx$1.1.

The high-redshift sources populate a specific area of the colour plane in both models, although 
the redshift distributions are different.
In the \citet{pearson07} model the highest redshift objects, z$>$3 occupy the parameter space 
corresponding to 
S$_{250}$/S$_{350}$ colours $<$1.0 with S$_{500}<$40mJy, 
while the 
\citet{xu01} model locates the z$>$3 region rather at 
 S$_{250}$/S$_{350}<$0.8 and the same flux cutoff, 
but with fewer objects and mixed with many low redshift SEDs.
Similar cuts can be made for z$>$2 sources. Lower redshift sources may also be excluded by 
virtue of their higher S$_{250}$/S$_{350}$ colours.

 The SPIRE data generally overlap fine at S$_{500}$$<$60, except for colours of S$_{250}$/S$_{350}<0.8$. 
Especially the \citet{pearson07} model shows no objects below this limit, while the 
same region is sparsely populated by the \citet{xu01} model SEDs. Looking at the model types, 
it turns out that those are mainly AGN, which are missing entirely in the \citet{pearson07} SED catalogue.
Neither model covers sufficiently the increasingly redder colours in this region 
that SPIRE observes towards 500~$\mu$m fluxes above 50~mJy. A comparison with another mock 
catalogue by \citet{valiante09} shows the same lack of red sources.

A considerable fraction of submm bright sources are expected to be 
lensed by foreground galaxies \citep{negrello07}. Since lensing magnification is 
wavelength independent, such lensed sources appear in their intrinsic positions in the 
colour-colour diagram, but their locations in the colour-flux plane would be offset to 
brighter fluxes (towards the right side of Fig.~\ref{Fig:colourflux} along the x-axis), 
while keeping colours the same. Using the models of \citet{negrello07} 
 and \citet{negrello10}
we estimate that 	
of all objects with fluxes S$_{500}>100$~mJy and redshift 2$<$z$<$3, almost all are lensed. 
For our 19.6 deg$^2$ total sky area, that would be $\approx$2-3 out of the 24 bright 500~$\mu$m sources 
we have identified,
although this is a lower limit, as our selection procedure may bias slightly against clustered and
as such potentially lensed sources.

For now we conclude that we see a population of red bright objects that may consist mostly
of colder SEDs but with a fraction of distant lensed ones. Inclusion of other wavelengths 
as shown by \citet{mrr10} will be needed for further interpretation.

\smallskip

\begin{acknowledgements}
SPIRE has been developed by a consortium of institutes led by Cardiff Univ. (UK) and 
including Univ. Lethbridge (Canada); NAOC (China); CEA, LAM (France); IFSI, Univ. Padua (Italy); 
IAC (Spain); Stockholm Observatory (Sweden); Imperial College London, RAL, UCL-MSSL, UKATC, 
Univ. Sussex (UK); and Caltech, JPL, NHSC, Univ. Colorado (USA). This development has been 
supported by national funding agencies: CSA (Canada); NAOC (China); CEA, CNES, CNRS (France); 
ASI (Italy); MCINN (Spain); SNSB (Sweden); STFC (UK); and NASA (USA). 
Support for this work was provided by NASA through an award issued by JPL, Caltech.
Data presented in this paper were analyzed using ÒThe {\it Herschel} Interactive Processing 
Environment (HIPE)Ó, a joint development by the {\it Herschel} Science Ground Segment 
Consortium, consisting of ESA, the NASA Herschel Science Center, and the HIFI, PACS and SPIRE consortia.
The data presented in this paper will be released through the {\it Herschel}  
Database in Marseille {\it HeDaM}\footnote{hedam.oamp.fr/HerMES}.
This work made substantial use of TOPCAT written by Mark Taylor\footnote{www.starlink.ac.uk/topcat}.
We thank Mattia Negrello for predictions of lensed counts.
Many thanks also to George Helou and an anonymous referee for helpful comments.
\end{acknowledgements}

\smallskip

\bibliographystyle{aa}

\end{document}